\documentstyle[twocolumn,aps]{revtex}
\begin{document}
\title{A single mode realaization of SU(1,1)}
\author{Vladimir L. Safonov\thanks{%
Email: vsafonov@ucsd.edu}}
\address{Center for Magnetic Recording Research, University of California -- San\\
Diego, 9500 Gilman Dr., La Jolla, CA 92093-0401, U.S.A.}
\date{\today}
\maketitle

\begin{abstract}
A single mode realization of SU(1,1) algebra is proposed. This realization
makes it possible to reduce the problem of two nonlinear interacting
oscillators to the problem of free moving particle.
\end{abstract}

\pacs{}

\section{Introduction}

The Lie algebra of SU(1,1) group is used in many branches of physics (see,
e.g., \cite{mattis}\cite{perelomov}\cite{mlodinov}\cite{gerry}\cite{gerry1} 
\cite{safonov}). These algebra consists of the three hyperbolic operators ${K%
}_{0}$, ${K}_{+}$, ${K}_{-}$\ which satisfy the following commutation
relations:

\begin{equation}
\lbrack {K}_{0},{K}_{\pm }]=\pm {K}_{\pm },\quad \lbrack {K}_{+},{K}_{-}]=-2{%
K}_{0}.  \label{Kcommutations}
\end{equation}
The Casimir invariant for this algebra is

\begin{equation}
{C}={K}_{0}^{2}-{\frac{1}{2}}({K}_{+}{K}_{-}+{K}_{-}{K}_{+}).
\label{casimir}
\end{equation}

The relations (\ref{Kcommutations}) and (\ref{casimir}) resemble,
respectively, the spin operator commutations

\begin{equation}
\lbrack {S}_{z},{S}_{\pm }]=\pm {S}_{\pm },\quad \lbrack {S}_{+},{S}_{-}]=2{S%
}_{z},  \label{scommut}
\end{equation}
where ${S}_{\pm }={S}_{x}\pm i{S}_{y}$, and the following invariant

\begin{equation}
S(S+1)=S_{z}^{2}+\frac{1}{2}(S_{+}S_{-}+S_{-}S_{+}).  \label{spin}
\end{equation}

This similarity can be continued. The SU(1,1) algebra has a single-bosonic
realization

\begin{eqnarray}
{K}_{-} &=&(2k+a^{+}a)^{1/2}a,  \nonumber \\
{K}_{+} &=&a^{+}(2k+a^{+}a)^{1/2},  \nonumber \\
{K}_{0} &=&k+a^{+}a,  \label{hopri} \\
{C} &=&k(k-1).  \nonumber
\end{eqnarray}
Here $a^{+}$ and $a$\ are the creation and annihilation bose operators,
respectively, with the commutation rule $[a,a^{+}]=1$. This representation
was considered first by Mlodinov and Papanicolaou \cite{mlodinov}, and it is
an analog of the Holstein-Primakoff \cite{holsteinprimakoff}\ representation
for spin operators:

\begin{eqnarray}
S_{-} &=&(2S-a^{+}a)^{1/2}a,  \nonumber \\
S_{+} &=&a^{+}(2S+a^{+}a)^{1/2},  \nonumber \\
S_{z} &=&-S+a^{+}a.  \label{holstein}
\end{eqnarray}

There is another important single-mode representation for spin proposed by
Villain \cite{villain}. In this Letter we shall find an analog of this
single mode representation for the hyperbolic operators and consider its
application.

\section{A single mode representation}

If $X$\ is a coordinate and $P=-id/dX$\ \ is the momentum operator, the
Villain representation can be written as

\begin{eqnarray}
S_{-} &=&\sqrt{(S+1/2)^{2}-(P-1/2)^{2}}\exp \left( -iX\right) ,  \nonumber \\
S_{+} &=&\exp \left( iX\right) \sqrt{(S+1/2)^{2}-(P-1/2)^{2}},  \nonumber \\
S_{z} &=&P.  \label{villain}
\end{eqnarray}

Due to the similarity between spin and hyperbolic operators mentioned above,
one can construct the following single mode representation:

\begin{eqnarray}
{K}_{-} &=&(P+P_{0})\exp \left( -iX\right) ,  \nonumber \\
{K}_{+} &=&\exp \left( iX\right) (P+P_{0}^{\ast }),  \nonumber \\
{K}_{0} &=&P+(P_{0}+P_{0}^{\ast })/2-1/2  \label{saf}
\end{eqnarray}
with

\begin{equation}
{C}=-1/4+(P_{0}-P_{0}^{\ast })^{2}/4.  \label{newcasimir}
\end{equation}
Here $P_{0}$\ is a complex number. Taking into account that $[X,P]=i$ and

\begin{equation}
\exp (i\beta X)P^{n}\exp (-i\beta X)=(P-\beta )^{n},  \label{transfo}
\end{equation}
where $\beta $ is a formal parameter and $n=1,2,...$ , it is simple to check
out the validity of commutations (\ref{Kcommutations}) for (\ref{saf}).

Another form of representaion (\ref{saf})\ may be obtained if we change $%
X\longrightarrow -P$ and $P\longrightarrow X$.

It should be noted that Perelomov \cite{perelomov}\ mentioned a single mode
realization of SU(1,1) in the form:

\begin{eqnarray}
{K}_{0} &=&-id/d\theta ,  \label{perelom} \\
{K}_{\pm } &=&-i\exp \left( \pm i\theta \right) d/d\theta \mp (-1/2+i\lambda
)\exp (\pm i\theta )  \nonumber
\end{eqnarray}
with

\begin{equation}
{C}=-1/4-\lambda ^{2}/4.\quad \lambda >0.  \label{perelcasimir}
\end{equation}
After simple algebra, one can find that (\ref{perelom}) can be transformed
to the form (\ref{saf}) with $\theta =X$ and $P_{0}=1/2+i\lambda $.

Substituting

\begin{equation}
Q=\frac{\alpha +\alpha ^{+}}{\sqrt{2}},\quad P=\frac{\alpha -\alpha ^{+}}{i%
\sqrt{2}},  \label{boseoperators}
\end{equation}
we obtain the following representations of the hyperbolic operators in terms
of bose operators $\alpha $ and $\alpha ^{+}$:

\begin{eqnarray}
{K}_{-} &=&\left( \frac{\alpha -\alpha ^{+}}{i\sqrt{2}}+P_{0}\right) \exp
\left( -i\frac{\alpha +\alpha ^{+}}{\sqrt{2}}\right) ,  \nonumber \\
{K}_{+} &=&\exp \left( i\frac{\alpha +\alpha ^{+}}{\sqrt{2}}\right) \left( 
\frac{\alpha -\alpha ^{+}}{i\sqrt{2}}+P_{0}^{\ast }\right) ,  \nonumber \\
{K}_{0} &=&\frac{\alpha -\alpha ^{+}}{i\sqrt{2}}+(P_{0}+P_{0}^{\ast })/2-1/2,
\label{form1}
\end{eqnarray}
or,

\begin{eqnarray}
{K}_{-} &=&\left( \frac{\alpha +\alpha ^{+}}{\sqrt{2}}+P_{0}\right) \exp
\left( \frac{\alpha -\alpha ^{+}}{\sqrt{2}}\right) ,  \nonumber \\
{K}_{+} &=&\exp \left( -\frac{\alpha -\alpha ^{+}}{\sqrt{2}}\right) \left( 
\frac{\alpha +\alpha ^{+}}{\sqrt{2}}+P_{0}^{\ast }\right) ,  \nonumber \\
{K}_{0} &=&\frac{\alpha +\alpha ^{+}}{\sqrt{2}}+(P_{0}+P_{0}^{\ast })/2-1/2.
\label{form2}
\end{eqnarray}
Note that a classical analog of the representation (\ref{form2}) has been
derived in \cite{safonov}.

\section{Example}

Let us consider now\ the following Hamiltonian:

\begin{eqnarray}
{\cal H} &=&\varepsilon (a^{+}a+b^{+}b)+\Phi _{2}a^{+}b^{+}ba  \nonumber \\
&&+\Phi _{1}(a^{+}a^{+}aa+b^{+}b^{+}bb).  \label{Hamexample}
\end{eqnarray}
It describes a system of two nonlinear interacting oscillators with the Bose
operators $a^{+}$, $a$ and $b^{+}$, $b$. The two mode representation of
SU(1,1) is defined by

\begin{eqnarray}
{K}_{-} &=&ab,\quad {K}_{+}=a^{+}b^{+},  \nonumber \\
{K}_{0} &=&\frac{1}{2}\left( a^{+}a+b^{+}b+1\right)  \label{twomode}
\end{eqnarray}
with

\begin{equation}
{C}={-1/4+(}a^{+}a-b^{+}b{)}^{2}{/4.}  \label{twocasimir}
\end{equation}

The Hamiltonian (\ref{Hamexample}) in terms of (\ref{twomode}) has the form:

\begin{eqnarray}
{\cal H} &=&2\Phi _{1}-\varepsilon +(2\varepsilon -6\Phi _{1})K_{0}
\label{Kform} \\
&&+4\Phi _{1}K_{0}^{2}+(\Phi _{2}-2\Phi _{1})K_{+}K_{-}.  \nonumber
\end{eqnarray}

Let us consider the pair states when the numbers of `$a$' and `$b$' bosons
are equal (formally, $a^{+}a=b^{+}b$). In this case the Casimir invariants (%
\ref{newcasimir}) and (\ref{twocasimir}) are equal to $-1/4$. Substituting (%
\ref{saf}) with 
\begin{equation}
P_{0}=(3\Phi _{1}+\Phi _{2}-\varepsilon )/(2\Phi _{1}+\Phi _{2})  \label{p0}
\end{equation}
into the (\ref{Kform}), we obtain the Hamiltonian

\begin{equation}
{\cal H}={\cal H}_{0}+P^{2}/2m.  \label{free}
\end{equation}
Here ${\cal H}_{0}=-(\Phi _{1}-\varepsilon )^{2}/(2\Phi _{1}+\Phi _{2})$ and 
$m=1/(4\Phi _{1}+2\Phi _{2})$. If $2\Phi _{1}+\Phi _{2}>0$, we have a ground
state (condensate) with the energy ${\cal H}_{0}$ and an excitation with a
quadratic spectrum of a particle of the mass $m$ and momentum $P$. Thus,
with the help of representation (\ref{saf}), we can reduce the nonlinear
Hamiltonian of two interacting oscillators (\ref{Hamexample}) to the
Hamiltonian (\ref{free}) of free moving particle.

\end{document}